\documentclass[a4paper,12pt]{article}
\usepackage{latexsym}
\usepackage{amssymb}
\usepackage{amsmath}
\usepackage{cite}
\usepackage{color}

\usepackage{microtype}
\usepackage{slashed}

\numberwithin{equation}{section}
\allowdisplaybreaks

\makeatletter \@addtoreset{equation}{section} \makeatother

\usepackage{cancel}

\usepackage{ulem}
\usepackage{ytableau}

\ytableausetup{centertableaux, mathmode, smalltableaux}

\definecolor{blue-violet}{rgb}{0.54, 0.17, 0.89}
\definecolor{PineGreen}{cmyk}{0.92, 0, 0.59, 0.25}
\definecolor{YellowOrange}{cmyk}{0, 0.42, 1, 0}

\newcommand{\be}{\begin{equation}}
\newcommand{\ee}{\end{equation}}
\newcommand{\beq} {\begin{equation}}
\newcommand{\eeq} {\end{equation}}
\newcommand{\ba}{\begin{eqnarray}}
\newcommand{\ea}{\end{eqnarray}}

\usepackage{hyperref}

\begin{document}
\numberwithin{equation}{section}


\begin{center}
{\bf\LARGE  Cosmology of Metric-Affine $R + \beta R^2$ Gravity with Pure Shear Hypermomentum} \\
\vskip 2 cm
{\bf \large Damianos Iosifidis$^{1,2}$, Ratbay Myrzakulov$^{3,4}$, Lucrezia Ravera$^{5,6}$}
\vskip 8mm
\end{center}
\noindent {\small $^{1}$ \it Laboratory of Theoretical Physics, Institute of Physics, University of Tartu, W. Ostwaldi 1, 50411 Tartu, Estonia. \\
$^{2}$ Institute of Theoretical Physics, Department of Physics, Aristotle University of Thessaloniki, 54124 Thessaloniki, Greece. \\
$^{3}$ \it Ratbay Myrzakulov Eurasian International Centre for Theoretical Physics, Nur-Sultan, 010009, Kazakhstan. \\
$^{4}$ \it Eurasian National University, Nur-Sultan, 010008, Kazakhstan. \\
$^{5}$ \it DISAT, Politecnico di Torino, Corso Duca degli Abruzzi 24, 10129 Torino, Italy. \\
$^{6}$  \it INFN, Sezione di Torino, Via P. Giuria 1, 10125 Torino, Italy.
}

\vskip 2 cm
\begin{center}
{\small {\bf Abstract}}
\end{center}

In this paper we study the cosmological aspects of metric-affine $f(R)$ gravity with hyperfluid. The equations of motion of the theory are obtained by varying the action with respect to the metric and the independent connection. Subsequently, considering a Friedmann-Lemaître-Robertson-Walker background, we derive the modified Friedmann equations in the presence of a perfect cosmological hyperfluid. Especially, we focus on the particular case in which $f(R)=R+\beta R^2$, considering purely shear hypermomentum and finding exact solutions in the weak coupling limit.

\vfill
\noindent {\small{\it
    E-mail:  \\
{\tt damianos.iosifidis@ut.ee;} \\
{\tt rmyrzakulov@gmail.com}; \\
{\tt lucrezia.ravera@polito.it}}}
   \eject

\tableofcontents

\noindent\hrulefill

\section{Introduction}

In past and recent years there has been a widely shared interest in gravitational theories beyond general relativity, especially under the cosmological perspective. Many alternative theories of gravity embrace a geometrical description of spacetime and are therefore based on modified geometrical scenarios, in particular on non-Riemannian geometry (see, e.g., \cite{Eisenhart:1927} and \cite{Klemm:2018bil} for a concise review). In this context, there emerges the rather general framework of metric-affine gravity (MAG) \cite{Hehl:1994ue,Hehl:1999sb,Sotiriou:2006qn,Capozziello:2007tj,Vitagliano:2010sr,Latorre:2017uve,Iosifidis:2018jwu,Hohmann:2019fvf,Iosifidis:2019jgi,BeltranJimenez:2019acz,Aoki:2019rvi,Percacci:2020ddy,BeltranJimenez:2020sqf,Bahamonde:2020fnq,Cabral:2020fax,Percacci:2020bzf,Rigouzzo:2022yan,Jimenez-Cano:2022sds,Baldazzi:2021kaf}, in which the metric and the connection are treated, a priori, as independent objects, without any assumptions on the general affine connection from the very beginning. The explicit form of the affine connection is
eventually obtained from the study of the field equations derived in the first order (i.e., Palatini) formalism. As a result, torsion and non-metricity are typically involved in MAG. Moreover, couplings of matter to the general affine connection are expressed by means of the so-called hypermomentum tensor \cite{Hehl:1976kt,Hehl:1976kt2,Ariwahjoedi:2021yth}, which describes dilation, spin, and shear, encompassing the microstructure of matter.

Several studies on cosmological aspects have been performed, especially in the last years, by considering the large class of MAG theories (see, e.g., \cite{Barragan:2009sq,BeltranJimenez:2015pnp,BeltranJimenez:2017vop,Iosifidis:2018diy,Kranas:2018jdc,Shimada:2018lnm,Bombacigno:2018tyw,Kubota:2020ehu,Mikura:2020qhc,Mikura:2021ldx,Iosifidis:2021crj,Bombacigno:2021bpk,Myrzakulov:2021vel,Iosifidis:2021iuw,Yang:2021fjy,Iosifidis:2021xdx,Papagiannopoulos:2022ohv,Anagnostopoulos:2020lec,Iosifidis:2021fnq,Gialamas:2022xtt}).\footnote{The literature on the subject is quite extended; here we reported the works that most inspired the analysis contained in the present paper.} Here we shall consider   $f(R)$ gravity in the metric-affine setup. It is known that Palatini $f(R)$ theories with matter (where both torsion and non-metricity can be involved, but the matter fields do not couple to the connection) are equivalent to a Brans-Dicke theory with Brans-Dicke parameter $\omega_{0}=-3/2$ (see \cite{Sotiriou:2006hs}). On the other hand, in metric-affine $f(R)$ theories, where the matter part of the action is allowed to contain couplings with the affine connection, there appear, in general, hypermomentum contributions to the connection field equations, which typically makes the study of such theories more involved under the computational perspective and no correspondence with Brans-Dicke theory exists under such conditions. However, the inclusion of hypermomentum is crucial to understand the interrelation between the microstructure of matter and extended geometry.

Modified gravity theories may also include curvature-squared corrections to the Einstein-Hilbert action. In particular, when the curvature is large, solving the Einstein's equations in the presence of curvature squared terms leads to an effective cosmological constant. In this context, in 1979 it was proposed that the early Universe went through an inflationary de Sitter era \cite{Starobinsky:1979ty,Starobinsky:1980te}, originally using the semi-classical Einstein's equations with free matter fields. Subsequently, it was realized that inflation can be controlled by a contribution from a squared Ricci scalar term in the effective action \cite{Vilenkin:1985md}, that is, in other words, by considering an $f(R)$ gravity theory such that $f(R)=R+\beta R^2$, where $\beta$ has  dimension of  inverse  mass squared. Correspondingly, the inflationary scenario associated to the emerging potential is commonly referred to as Starobinsky inflation.

In the present work we extend the analysis of this cosmological model to the metric-affine framework, in the presence of a perfect hyperfluid (which is a classical continuous medium carrying hypermomentum, see, e.g., \cite{Obukhov:1993pt,Obukhov:1996mg,Babourova:1998mgh}). In particular, we start from the study of metric-affine $f(R)$ gravity, deriving the field equations of the theory and the modified Friedmann equations in a Friedmann-Lemaître-Robertson-Walker (FLRW) background, in the presence of a perfect cosmological hyperfluid. Then, we focus on the specific   $f(R)=R+\beta R^2$ theory, which we analyze thoroughly.

The paper is organized as follows: In Section \ref{tb} we give the geometrical and theoretical background. In Section \ref{thethy} we derive the field equations and the modified Friedmann equations of metric-affine $f(R)$ gravity considering a FLRW background in the presence of a perfect cosmological hyperfluid. Consequently, in Section \ref{speccase} we focus on the cosmology of the $R+\beta R^2$ metric-affine theory, obtaining exact solutions in the weak coupling limit (i.e., $\beta  R <<1$) in the case of purely shear hypermomentum. Section \eqref{conclusions} is devoted to some final remarks.

\section{Theoretical background}\label{tb}

Let us now start by briefly introducing the basic geometrical aspects along with the necessary theoretical background needed for the rest of our analysis.

\subsection{Geometric setup}

We work in the framework of non-Riemannian geometry, where we have a metric tensor $g_{\mu \nu}$ (we will consider four spacetime dimensions, that is $\mu, \nu, \ldots=0,1,2,3$ and a mostly plus signature) and a general affine connection ${\Gamma^\lambda}_{\mu \nu}$,\footnote{These two objects will be considered, a priori, as independent.} whose generic decomposition is
\begin{equation}\label{decgamma}
{\Gamma^\lambda}_{\mu \nu} = \tilde{\Gamma}^\lambda_{\phantom{\lambda} \mu \nu} + {N^\lambda}_{\mu \nu}\,,
\end{equation}
where
\beq\label{lcconn}
\tilde{\Gamma}^\lambda_{\phantom{\lambda}\mu\nu} = \frac12 g^{\rho\lambda}\left(\partial_\mu 
g_{\nu\rho} + \partial_\nu g_{\rho\mu} - \partial_\rho g_{\mu\nu}\right)
\eeq
is the Levi-Civita connection and the tensor ${N^\lambda}_{\mu\nu}$ is given in terms of torsion
\beq
{S_{\mu\nu}}^\lambda := {\Gamma^\lambda}_{[\mu\nu]}\,, \quad S_{\mu \nu \alpha} = N_{\alpha[\mu \nu]} \label{torsdef}
\eeq
and non-metricity
\beq
Q_{\lambda\mu\nu}:= -\nabla_\lambda g_{\mu\nu} = 
-\partial_\lambda g_{\mu\nu} + {\Gamma^\rho}_{\mu\lambda} g_{\rho\nu} +
{\Gamma^\rho}_{\nu\lambda}g_{\mu\rho} \,, \quad Q_{\nu \alpha \mu} = 2 N_{(\alpha \mu )\nu} \label{nonmetdef}
\eeq
as follows:
\beq\label{distortion}
{N^\lambda}_{\mu\nu} = {\frac12 g^{\rho\lambda}\left(Q_{\mu\nu\rho} + Q_{\nu\rho\mu}
- Q_{\rho\mu\nu}\right)} - {g^{\rho\lambda}\left(S_{\rho\mu\nu} +
S_{\rho\nu\mu} - S_{\mu\nu\rho}\right)} \,.
\eeq
We can write the following trace decomposition for the torsion and non-metricity tensors, respectively (holding in four spacetime dimensions) \cite{Hehl:1994ue,Iosifidis:2019jgi}:
\begin{equation}\label{dectorandnm}
\begin{split}
{S_{\lambda\mu}}^\nu & = \frac{2}{3} {\delta_{[\mu}}^{\nu} S_{\lambda]} + \frac{1}{6} \varepsilon_{\lambda \mu \kappa \rho} g^{\kappa \nu} t^\rho + {Z_{\lambda\mu}}^\nu \,,  \\
Q_{\lambda\mu\nu} & = \frac{5}{18} Q_\lambda g_{\mu\nu} - \frac19 q_\lambda g_{\mu\nu} + \frac49 g_{\lambda(\nu}q_{\mu)} - \frac19 g_{\lambda(\nu} Q_{\mu)} + \Omega_{\lambda\mu\nu} \,, 
\end{split}
\end{equation}
where $Q_\lambda := {Q_{\lambda \mu}}^\mu$ and $q_\nu := {Q^\mu}_{\mu\nu}$ are the non-metricity vectors, $S_\lambda :={S_{\lambda \sigma}}^{\sigma}$ is the torsion vector, $t^\rho := \varepsilon^{\rho \lambda \mu \nu} S_{\lambda \mu \nu}$ is the torsion pseudo-vector, ${Z_{\lambda\mu}}^\nu$ is the traceless part of torion, and $\Omega_{\lambda\mu\nu}$ is the traceless part of non-metricity. \\
We define the curvature (Riemann) tensor as
\beq\label{curvtensdef}
{R^\mu}_{\nu \alpha \beta} := 2 \partial_{[\alpha} {\Gamma^\mu}_{|\nu|\beta]} + 2 {\Gamma^\mu}_{\rho[\alpha} {\Gamma^\rho}_{|\nu|\beta]} = \tilde{R}^\mu_{\phantom{\mu} \nu \alpha \beta} + 2 \tilde{\nabla}_{[\alpha} {N^\mu}_{|\nu|\beta]} + 2 {N^\mu}_{\lambda|\alpha} {N^\lambda}_{|\nu|\beta]} \,,
\eeq
where $\tilde{\nabla}$ denotes the Levi-Civita covariant derivative and $\tilde{R}^\mu_{\phantom{\mu} \nu \alpha \beta}$ is the associated Riemann tensor. The Ricci tensor of $\Gamma$ is $R_{\nu \beta} := {R^\mu}_{\nu \mu \beta}$ and the associated curvature scalar is $R:= R_{\mu \nu} g^{\mu \nu}$.

\subsection{Energy-momentum tensors and hypermomentum}

Let us now recall the concepts of energy-momentum and hypermomentum tensors, following \cite{Iosifidis:2020gth}. We assume the full action to be a functional of the metric (and its derivatives), the general affine connection, and the matter fields (denoted by $\varphi$), that is
\beq \label{magact}
S[g,\Gamma,\varphi] = S_{\text{G}}[g,\Gamma] + S_{\text{M}}[g,\Gamma,\varphi] \,,
\eeq
where the gravitational and matter part of the action can be respectively written as
\beq
S_{\text{G}}[g,\Gamma] = \frac{1}{2\kappa} \int d^n x \sqrt{-g} \mathcal{L}_{\text{G}} (g,\Gamma) \,, \quad S_{\text{M}}[g,\Gamma,\varphi] = \int d^n x \sqrt{-g} \mathcal{L}_{\text{M}} (g,\Gamma,\varphi) \,,
\eeq
with $\kappa=8\pi G$ the gravitational constant.
One can then define the metric energy-momentum tensor (MEMT),
\beq
T_{\mu \nu} := - \frac{2}{\sqrt{-g}} \frac{\delta S_{\text{M}}}{\delta g^{\mu \nu}} = - \frac{2}{\sqrt{-g}} \frac{\delta (\sqrt{-g} \mathcal{L}_{\text{M}})}{\delta g^{\mu \nu}} \,,
\eeq
and the hypermomentum tensor \cite{Hehl:1994ue,Hehl:1976kt,Hehl:1976kt2},
\beq
{\Delta_\lambda}^{\mu \nu} := - \frac{2}{\sqrt{-g}} \frac{\delta S_{\text{M}}}{\delta {\Gamma^\lambda}_{\mu \nu}} = - \frac{2}{\sqrt{-g}} \frac{\delta (\sqrt{-g} \mathcal{L}_{\text{M}})}{\delta {\Gamma^\lambda}_{\mu \nu}} \,.
\eeq
Working in the equivalent formalism based on the vielbein ${e_{\mu}}^c$ and spin connection $\omega_{\mu | a b}$, where $a,b,\ldots$ are Lorentz (i.e., tangent space) indices, one may also introduce the so-called canonical energy-momentum tensor (CEMT),
\beq\label{cemt}
{t^\mu}_c := \frac{1}{\sqrt{-g}} \frac{\delta S_{\text{M}}}{\delta {e_\mu}^c} \,.
\eeq
The following relation holds \cite{Hehl:1994ue,Iosifidis:2020gth}:
\beq\label{cemt1}
{t^\mu}_\lambda := {T^\mu}_\lambda - \frac{1}{2 \sqrt{-g}} \hat{\nabla}_\nu \left( \sqrt{-g} {\Delta_\lambda}^{\mu \nu} \right) \,,
\eeq
where
\beq\label{hatnabla}
\hat{\nabla}_\nu := 2 S_\nu - \nabla_\nu \,,
\eeq
which implies
\beq\label{tracerelation}
t = T + \frac{1}{2 \sqrt{-g}} \partial_\nu \left( \sqrt{-g} \Delta^\nu \right) \,,
\eeq
with
\beq
t := {t^\mu}_\mu \,, \quad T := {T^\mu}_\mu \,, \quad \Delta^\nu := {\Delta_\lambda}^{\lambda \nu} \,.
\eeq
Let us conclude by saying that, in four spacetime dimensions, the hypermomentum tensor can be decomposed as follows \cite{Iosifidis:2020upr}:
\begin{equation}
\Delta_{\alpha \mu \nu} = \tilde{\Delta}_{\alpha \mu \nu} + \frac{1}{4} g_{\alpha \mu} D_\nu + \mathring{\Delta}_{\alpha \mu \nu} \,,
\end{equation}
where $\tilde{\Delta}_{\alpha \mu \nu} := \Delta_{[\alpha \mu] \nu}$ is the spin part, $D^\nu := {\Delta_\mu}^{\mu \nu}$ is the dilation, and $\mathring{\Delta}_{\alpha \mu \nu} := \Delta_{(\alpha \mu)_0 \nu}$ the shear, that is traceless and symmetric in the indices $\alpha,\mu$.

\subsection{Non-Riemannian FLRW cosmology}\label{cosmtornonmetsubsec}

In the following we recall key cosmological aspects in the framework of non-Riemannian geometry, which will be useful in the reminder of the paper. \\
First of all, we will consider a homogeneous and isotropic, flat FLRW spacetime with the usual Robertson-Walker line element
\beq\label{RWle}
ds^{2}=-dt^{2}+a^{2}\delta_{ij}dx^{i}dx^{j} \,,
\eeq
where $a(t)$ is the scale factor of the Universe and $i,j=1,2,3$. We then define the projector tensor
\beq\label{projop}
h_{\mu \nu}:= g_{\mu \nu} + u_\mu u_\nu \,,
\eeq
where $u^\mu$ is the normalized $n$-velocity field of a given fluid which, in co-moving coordinates, is expressed as $u^\mu= \delta^\mu_0=(1,0,0,0)$, $u_\mu u^\mu=-1$. Accordingly, we introduce the temporal derivative
\beq\label{tempdev}
\dot{}=u^\alpha \nabla_\alpha \,.
\eeq
The above constitutes a $1+3$ spacetime split. \\
In a non-Riemannian FLRW spacetime in $1+3$ dimensions the general affine connection can be written as \cite{Iosifidis:2020gth}
\begin{equation}\label{connFLRW}
{\Gamma^\lambda}_{\mu \nu} = \tilde{\Gamma}^\lambda_{\phantom{\lambda}\mu \nu} + X(t) u^\lambda h_{\mu \nu} + Y(t) u_\mu {h^\lambda}_\nu + Z(t) u_\nu {h^\lambda}_\mu + V(t) u^\lambda u_\mu u_\nu + {\varepsilon^\lambda}_{\mu \nu \rho} u^\rho W(t) \,,
\end{equation}
while the torsion and non-metricity tensors can be written, respectively, in the following way \cite{Iosifidis:2020gth}:\footnote{The fact that isotropic and homogeneous torsion has 2 components was proven in \cite{Tsamparlis:1979} and that non-metricity has 3 respectively was shown in \cite{Minkevich:1998cv}.}
\begin{equation}\label{tornonmetFLRW}
\begin{split}
S^{(n)}_{\mu \nu \alpha} & = 2 u_{[\mu} h_{\nu]\alpha} \Phi(t) + \varepsilon_{\mu \nu \alpha \rho} u^\rho P(t) \,, \\
Q_{\alpha \mu \nu} & = A(t) u_\alpha h_{\mu \nu} + B(t) h_{\alpha(\mu} u_{\nu)} + C(t) u_\alpha u_\mu u_\nu \,.
\end{split}
\end{equation}
The functions $X(t)$, $Y(t)$, $Z(t)$, $V(t)$, $W(t)$ in \eqref{connFLRW} and $\Phi(t)$, $P(t)$, $A(t)$, $B(t)$, $C(t)$ in \eqref{tornonmetFLRW} describe non-Riemannian cosmological effects. \\
Using the decomposition of $\Gamma$, one can then prove that
\begin{equation}\label{connvarintermsoftnm}
W = P \,, \quad V= \frac{C}{2} \,, \quad Z = \frac{A}{2} \,, \quad Y = 2 \Phi + \frac{A}{2} \,, \quad X = \frac{B}{2} - 2 \Phi - \frac{A}{2} \,.
\end{equation}
These are key ingredients to derive the modified Friedmann equations.

\subsection{Perfect cosmological hyperfluid}\label{perfhypsubs}

The general formulation of perfect cosmological hyperfluid generalizing the classical perfect fluid notion can be found in \cite{Iosifidis:2020gth,Iosifidis:2021nra}. We will consider a perfect cosmological hyperfluid in a homogeneous cosmological setting, demanding also isotropy. \\
The perfect hyperfluid is described in terms of the following MEMT and CEMT tensors \cite{Iosifidis:2020gth}:
\begin{align}
    & T_{\mu \nu} = \rho u_\mu u_\nu + p h_{\mu \nu} \,, \label{metrenmomform} \\
    & t_{\mu \nu} = \rho_c u_\mu u_\nu + p_c h_{\mu \nu} \,, \label{canonenmomform}
\end{align}
where $\rho$ and $p$ are the usual density and pressure of the perfect fluid component of the hyperfluid, while $\rho_c$ and $p_c$ are, respectively, the canonical (net) density and canonical pressure of the hyperfluid. Besides, the hypermomentum tensor associated with the perfect hyperfluid is
\beq\label{hypermomform}
\Delta^{(n)}_{\alpha \mu \nu} = \phi(t) h_{\mu \alpha} u_\nu + \chi(t) h_{\nu \alpha} u_{\mu} + \psi(t) u_{\alpha} h_{\mu \nu} + \omega(t) u_\alpha u_\mu u_\nu + \delta^n_4 \varepsilon_{\alpha \mu \nu \rho} u^\rho \zeta(t) \,.
\eeq
In general, one has the following conservation laws \cite{Iosifidis:2020gth} (see also \cite{Obukhov:2014nja}):
\begin{align}
& \frac{1}{\sqrt{-g}} \hat{\nabla}_\mu \left( \sqrt{-g} {t^\mu}_\alpha \right) = \frac{1}{2} \Delta^{\lambda \mu \nu} R_{\lambda \mu \nu \alpha} + \frac{1}{2} Q_{\alpha \mu \nu} T^{\mu \nu} + 2 S_{\alpha \mu \nu} t^{\mu \nu} \,, \label{conslawshyp1} \\
& {t^\mu}_\lambda = {T^\mu}_\lambda - \frac{1}{2 \sqrt{-g}} \hat{\nabla}_\nu \left( \sqrt{-g} {\Delta_\lambda}^{\mu \nu} \right) \,. \label{conslawshyp2}
\end{align}
Observe that \eqref{conslawshyp2} coincides with \eqref{cemt1}. One can then use the latter of the above equations in order to eliminate $t^{\mu\nu}$ from the former, yielding a variant conservation law  
\beq
	\sqrt{-g}(2 \tilde{\nabla}_{\mu}T^{\mu}_{\;\;\alpha}-\Delta^{\lambda\mu\nu}R_{\lambda\mu\nu\alpha})+\hat{\nabla}_{\mu}\hat{\nabla}_{\nu}(\sqrt{-g}\Delta_{\alpha}^{\;\;\mu\nu})+2S_{\mu\alpha}^{\;\;\;\;\lambda}\hat{\nabla}_{\nu}(\sqrt{-g}\Delta_{\lambda}^{\;\;\;\mu\nu})=0 \,.
\eeq
Let us conclude by mentioning that, given the most general form \eqref{hypermomform} of hypermomentum compatible with the cosmological principle, its spin, dilation, and shear parts read, respectively,
\begin{align}
& \tilde{\Delta}_{\alpha \mu \nu} = \left( \psi - \chi \right) u_{[\alpha} h_{\mu]\nu} + \epsilon_{\alpha \mu \nu \rho} u^\rho \zeta \,, \label{hspin} \\
& D_\nu := \Delta_{\alpha \mu \nu} g^{\alpha \mu} = \left( 3 \phi - \omega \right) u_\nu \,, \label{hdilation} \\
& \mathring{\Delta}_{\alpha \mu \nu} = \Delta_{(\alpha \mu )\nu} - \frac{1}{4} g_{\alpha \mu} D_\nu = \frac{\left( \phi + \omega \right)}{4} \left( h_{\alpha \mu} + 3 u_\alpha u_\mu \right) u_\nu + \left( \psi + \chi \right) u_{(\mu} h_{\alpha ) \nu} \,, \label{hshear}
\end{align}
in terms of the cosmological variables previously introduced.

\section{Metric-affine $f(R)$ gravity theory with hyperfluid}\label{thethy}

Let us now consider the action
\begin{equation}\label{actionfRhyp}
    S = \frac{1}{2 \kappa} \int d^4 x \sqrt{-g} f(R) + S_{\text{hyp.}} \,,
\end{equation}
where $f(R)$ is an arbitrary function of the scalar curvature $R:= g^{\mu \nu}R_{\mu \nu}(\Gamma)$, with $\Gamma$ a general affine connection, and $S_{\text{hyp.}}$ the hyperfluid action. 
Varying this action with respect to $g^{\mu \nu}$ we get
\begin{equation}\label{eomg}
    f'(R) R_{(\mu \nu)} - \frac{f(R)}{2} g_{\mu \nu} = \kappa T_{\mu \nu} \,.
\end{equation}
Taking the trace of this equation we obtain
\begin{equation}
    f'(R) R - 2 f(R) = \kappa T \,, \label{tracemet}
\end{equation}
that is
\begin{equation}
    {\frac{f}{f'}} = \frac{R}{2} - \frac{\kappa}{2f'} T \,,
\end{equation}
where $f'=f'(R)$. Plugging this expression back into \eqref{eomg}, the latter becomes
\begin{equation}
    R_{(\mu \nu)} - \frac{1}{4}  g_{\mu \nu} R = \frac{\kappa}{f'} \mathring{T}_{\mu \nu} \,,
\end{equation}
where $\mathring{T}_{\mu \nu}$ is the traceless part of the energy-momentum tensor, and it is defined as
\begin{equation}
    \mathring{T}_{\mu \nu} := T_{\mu \nu} - \frac{1}{4} g_{\mu \nu} T \,.
\end{equation}
On the other hand, varying the action with respect to the general affine connection ${\Gamma^\lambda}_{\mu \nu}$ we get the field equations
\beq
{P_{\lambda}}^{\mu\nu} + {\delta_\lambda}^\nu g^{\mu \sigma} \frac{\partial_\sigma f'}{f'} - g^{\mu \nu} \frac{\partial_\lambda f'}{f'} = \frac{\kappa}{f'} {\Delta_{\lambda}}^{\mu \nu} \,, \label{con}
\eeq
where ${P_{\lambda}}^{\mu\nu}$ is the Palatini tensor (which is traceless in the indices $\mu,\lambda$), defined as
\begin{equation}\label{palatinidefin}
{P_{\lambda}}^{\mu\nu} := -\frac{\nabla_{\lambda}(\sqrt{-g}g^{\mu\nu})}{\sqrt{-g}}+\frac{\nabla_{\sigma}(\sqrt{-g}g^{\mu\sigma})\delta^{\nu}_{\lambda}}{\sqrt{-g}} +2(S_{\lambda}g^{\mu\nu}-S^{\mu}\delta_{\lambda}^{\nu}+g^{\mu\sigma}{S_{\sigma\lambda}}^{\nu}) \,.
\end{equation}
Taking the different traces of \eqref{con}, along with other manipulations, the field equations of the connection yield the following set of equations:
\begin{align}
    & {\Delta^\rho}_{\rho \mu} = D_\mu = 0 \,, \label{dilzero} \\
    & S_\mu = \frac{3}{4} \left( \partial_\mu \ln f' - q_\mu \right) + \frac{\kappa}{8 f'} \left( {\Delta^\rho}_{\mu \rho} + 3 {\Delta_{\mu \rho}}^\rho \right) \,, \\
    & Q_\mu = 4 q_\mu - \frac{\kappa}{f'} \left( {\Delta^\rho}_{\mu \rho} + {\Delta_{\mu \rho}}^\rho \right) \,, \label{Qq} \\
    & t_\mu = - \frac{\kappa}{2 f'} \varepsilon_{\mu \nu \rho \sigma} {\Delta}^{\nu \rho \sigma} \,,
\end{align}
together with the fact that $\Omega_{\lambda \mu \nu}$ and $Z_{\lambda \mu \nu}$ result to be completely expressed in terms of the hypermomentum tensor (and $f'$). Notice that, in particular, \eqref{dilzero} means that the dilation part of the hypermomentum tensor vanishes.
The final form of the affine connection results to be
\begin{equation}
\begin{split}
    {\Gamma^\lambda}_{\mu \nu} & = \tilde{\Gamma}^\lambda_{\phantom{\lambda} \mu \nu} + \frac{\kappa}{f'} \frac{g^{\lambda \alpha}}{2} \left( \Delta_{\alpha \mu \nu} - \Delta_{\nu \alpha \mu} - \Delta_{\mu \nu \alpha} \right) + \frac{\kappa}{f'} \frac{g^{\alpha \lambda}}{2} g_{\nu[\mu} \left( \Delta_{\alpha]} - \tilde{\Delta}_{\alpha]} \right) \\
    & + \frac{1}{2f'} \left( \delta_\nu^\lambda \partial_\mu f' - g_{\mu \nu} \delta^\lambda f' \right) \,, 
\end{split}
\end{equation}
where we have defined $\Delta_\alpha:={\Delta_{\lambda \alpha}}^\lambda$ and $\tilde{\Delta}_\alpha:=g_{\lambda \mu}{\Delta_\alpha}^{\lambda \mu}={\Delta_{\alpha \lambda}}^\lambda$, and exploited the projective invariance (see, e.g., \cite{Iosifidis:2018zwo}) to remove the contribution in $q_\nu$ (that is, $\frac{1}{2} \delta_\mu^\lambda q_\nu$). 
Of course the vanishing of the dilation component is expected since the gravitational part of the action (i.e., $f(R)$) is invariant under projective transformations of the connection. Now, the trace equation \eqref{tracemet} implicitly defines the function $R=R(T)$ which, then, implies that $f(R)=f(R(T))=f(T)$.

\subsection{Cosmology of metric-affine $f(R)$ gravity with hyperfluid}\label{cosmMAGfRhyp}

With this in mind, and using the cosmological ansatz, from the connection field equations \eqref{con} we easily extract the relations
\begin{align}
    & \frac{A}{2}+4 \Phi-\frac{C}{2}=\frac{1}{F}\Big( \kappa \psi -\dot{F} \Big) \,, \\
    & B-\frac{3}{2}A-4 \Phi -\frac{C}{2}=\frac{1}{F}\Big( \kappa \chi +\dot{F} \Big) \,, \\
    & B=-2\frac{\kappa \phi}{F}\,, \\
    & B=-\frac{2}{3}\frac{\kappa \omega}{F} \,, \\
    &
    P=-\frac{1}{2}\frac{\kappa \zeta}{F} \,,
\end{align}
where we have set $F=f'$. Notice that, due to the vanishing of dilation, we have
\begin{equation}
    \omega=3 \phi 
\end{equation}
and, therefore, the two expressions for $B$ above are basically a single relation. \\
Next, since we have projective invariance, we can always set the gauge in such a way to ensure that one vectorial degree of freedom vanishes. Picking the gauge for which $q_{\mu}=0$, we get the extra relation
\begin{equation}
    C=\frac{3}{2}B \,. \label{C}
\end{equation}              
With this we can then solve the system above for torsion and non-metricity in terms of the sources, obtaining
\begin{align}
    & B=-2 \frac{\kappa \phi}{F}\,, \label{Beq} \\ 
    & C=-3\frac{\kappa \phi}{F}\,, \\
    & P=-\frac{1}{2}\frac{\kappa \zeta}{F} \,, \\
    & A=\frac{\kappa}{F}\Big[ \phi -(\chi+\psi) \Big] \,, \\
    & \Phi=\frac{\kappa}{4 F}\Big[ \frac{(\chi+3\psi)}{2}-2\phi \Big] -\frac{\dot{F}}{4 F} \,. \label{Phieq}
\end{align}
In addition, we see that the torsion function $\Phi$ contains also derivative terms of the energy-momentum trace, as they appear in $\dot{F}/F$. This means that torsion is excited even in the absence of hypermomentum, while non-metricity vanishes. This is clearly a consequence of the gauge choice we made. We could just as well have made a gauge choice of zero torsion vector, which then would imply a non-vanishing non-metricity even when the hypermomentum sources were switched-off. \\
Let us now write down the Friedmann equations for the general metric-affine $f(R)$ case. Firstly, we start with the acceleration equation. Its generic form for metric-affine spaces has been obtained in 
\cite{Iosifidis:2020zzp} and reads
\begin{align}
\frac{\ddot{a}}{a} &=-\frac{1}{3}R_{\mu\nu}u^{\mu}u^{\nu}+2\left( \frac{\dot{a}}{a} \right)\Phi +2\dot{\Phi} \nonumber \\
&+\left( \frac{\dot{a}}{a} \right)\left(A+\frac{C}{2}\right) +\frac{\dot{A}}{2}-\frac{A^{2}}{4}-\frac{1}{4}AC 
-A\Phi-C \Phi \,. \label{acceleq}
\end{align}
Then, contracting the metric field equations with $u^{\mu}u^{\nu}$, using also the above expressions of $A$, $B$, $C$, $\Phi$, and $P$ in terms of the hypermomentum sources, and recalling that the energy-momentum tensor has the usual perfect fluid form, we finally arrive at
\begin{align}
\frac{\ddot{a}}{a} &=-\frac{\kappa \rho}{3 F}+\frac{f}{6 F}-\frac{1}{2}\frac{\ddot{F}}{F}+\frac{1}{2}\left( \frac{\dot{F}}{F}\right)^{2}-\frac{\kappa \dot{F}}{2 F^{2}} {\psi}+\frac{\kappa}{4 F}(\dot{\psi}-\dot{\chi}
-2 \dot{\phi})\nonumber \\
&-\frac{\kappa}{4 F}\frac{\dot{a}}{a}(6 \phi+3 \chi+\psi)-\frac{1}{2}\frac{\dot{F}}{F}\frac{\dot{a}}{a}+\frac{\kappa^{2}}{8 F^{2}}\Big[ \psi^{2}-(2 \phi+\chi)^{2}\Big] \,,
\end{align}
where $F$ and $f$ are understood as functions of $T$ once the trace equation \eqref{tracemet} is solved. It is also worth noting the apparent similarity of the derivative terms for $a$ and $F$. \\
On the other hand, to derive the (modified) first Friedmann equation, let us notice that eq. \eqref{tracemet} can be rewritten as
\begin{equation}
    R = \frac{1}{f'} \left( 2 f + \kappa T \right) \,.
\end{equation}
Expanding the left-hand side of the latter by using the FLRW decomposition of the general affine connection and the fact that, from \eqref{metrenmomform}, we have $T=-\rho +3 p$, we get
\begin{equation}
    3\Big[(\dot{X}-\dot{Y})+3 H (X-Y) +(X+Y)(Z+V)-2 X Y - 2 W^{2}+2 \dot{H}+4 H^{2} \Big]= \frac{1}{F} \Big[ 2 f -\kappa (\rho -3 p) \Big] \,, \label{generaltreq}
\end{equation}
where we recall that $H:=\frac{\dot{a}}{a}$. The final form of the modified first Friedmann equation is then simply obtained by using the acceleration equation to eliminate the term $\dot{H}=\frac{\ddot{a}}{a}-H^2$ from eq. \eqref{generaltreq} and by plugging \eqref{Beq}-\eqref{Phieq} into \eqref{connvarintermsoftnm}. It reads as follows:
\begin{align}
H^{2} & =-\frac{\kappa}{{6F}}(\rho- 3 p)+\frac{\kappa \rho}{3 F}+\frac{f}{6 F}-\frac{1}{4}\left( \frac{\dot{F}}{F}\right)^{2}+\frac{\kappa \dot{F}}{2 F^{2}} {\psi}+\frac{\kappa}{2 F} \left(\dot{\phi} + 3 H \phi \right) \nonumber \\
& + H \left( \frac{\kappa}{4F} \psi - \frac{\dot{F}}{F} \right) + \frac{\kappa^{2}}{8 F^{2}} \Bigg[ \frac{1}{2} \chi^2 + 2 \chi \phi + 2 \phi^2 + \chi \psi - 2 \phi \psi - \frac{3}{2} \psi^2 - 4 \zeta ^2 \Bigg] \,, \label{fffin1}
\end{align}
where, as we will discuss below, the expression of $\dot{\phi}$ is then given by a conservation law of the perfect cosmological hyperfluid.
Notice that \eqref{fffin1} can also be rewritten as
\begin{align}
    \left( H + \frac{1}{2} \frac{\dot{F}}{F} \right)^2 & =-\frac{\kappa}{{6F}}(\rho- 3 p)+\frac{\kappa \rho}{3 F}+\frac{f}{6 F}+\frac{\kappa \dot{F}}{2 F^{2}} {\psi}+\frac{\kappa}{2 F} \left(\dot{\phi} + 3 H \phi \right) \nonumber \\
    & + H \frac{\kappa}{4F} \psi + \frac{\kappa^{2}}{8 F^{2}} \Bigg[ \frac{1}{2} \chi^2 + 2 \chi \phi + 2 \phi^2 + \chi \psi - 2 \phi \psi - \frac{3}{2} \psi^2 - 4 \zeta ^2 \Bigg] \,,
\end{align}
with a perfect square on the left-hand side. It is worth stressing out that the double derivative terms $\ddot{F}$ have canceled out and are absent from the 1st Friedmann equation. \\
The above were derived for $T\neq 0$. For conformally invariant matter (i.e., $T=0$) on-shell the trace equation \eqref{tracemet} would have a number of solutions $R =R_{0}=\text{constant}$ and subsequently $f(R)=f(R_{0})=f_{0}=\text{constant}$ as well as $F(R)=F(R_{0})=F_{0}=\text{constant}$. In this instance the Friedmann equations become
\begin{align}
    H^{2}&=\frac{\kappa \rho}{3F_0}+\frac{f_{0}}{6 F_{0}}+\frac{\kappa}{2 F_{0}}(\dot{\phi}+3 H \phi) 
    +\frac{\kappa}{4 F_{0}}H\psi \nonumber \\
    &+\frac{\kappa^{2}}{16 F_{0}^{2}}\Big( (\chi+2 \phi)^{2}+ 2\chi \psi -4 \phi \psi -3\psi^{2}-8 \zeta^{2}\Big)
\end{align}
and 
\begin{align}
\frac{\ddot{a}}{a} &=-\frac{\kappa \rho}{3 F_{0}}+\frac{f_{0}}{6 F_{0}}+\frac{\kappa}{4 F_{0}}(\dot{\psi}-\dot{\chi}
-2 \dot{\phi})\nonumber \\
&-\frac{\kappa}{4 F_{0}}\frac{\dot{a}}{a}(6 \phi+3 \chi+\psi)+\frac{\kappa^{2}}{8 F_{0}^{2}}\Big[ \psi^{2}-(\chi+2 \phi)^{2}\Big] \,.
\end{align}
It is worth stressing out that in this case there are no coupling terms between the hypermomentum current and the perfect fluid contributions. In addition, the modifications come now only from hypermomentum compared to the classical case. Furthermore, if there exists the $R=0$ solution and given that $f(R)$ is analytic on an open disk around $R=0$, namely the Taylor series\footnote{Note that here we are considering the sum starting from $n=1$, namely we do not include the constant term $C_{0}$ since this would correspond to a cosmological constant. Of course this inclusion is by all means possible but outside of the scope of the current study.}
\beq
f(R)=\sum_{n=1}^{\infty}C_{n}R^{n}
\eeq
exists and converges for all $R$ in this disk, then on-shell
\beq
f_{0}=f(0)=0 \,, \quad
F_{0}=F(0)=1 \,,
\eeq
where the value $C_{1}=1$ has been assumed to guarantee the proper general relativity limit. Under such circumstances we get further simplifications and the above Friedmann equations become
\begin{gather}
    H^{2}=\frac{\kappa \rho}{3}+\frac{\kappa}{2 }(\dot{\phi}+3 H \phi) 
    +\frac{\kappa}{4}H\psi 
    +\frac{\kappa^{2}}{16}\Big( (\chi+2 \phi)^{2}+ 2\chi \psi -4 \phi \psi -3\psi^{2}-8 \zeta^{2}\Big)
\end{gather}
and 
\begin{align}
\frac{\ddot{a}}{a} &=-\frac{\kappa \rho}{3}+\frac{\kappa}{4}(\dot{\psi}-\dot{\chi}
-2 \dot{\phi})
-\frac{\kappa}{4}\frac{\dot{a}}{a}(6 \phi+3 \chi+\psi)+\frac{\kappa^{2}}{8}\Big[ \psi^{2}-(\chi+2 \phi)^{2}\Big] \,.
\end{align}
We shall now proceed with an in-depth analysis of the particular case in which $f(R)=R+\beta R^2$.

\section{Special case $f(R)=R+\beta R^2$}\label{speccase}

Let us now consider the special case in which
\begin{equation}
    f(R)=R+\beta R^2 \,,
\end{equation}
where $\beta$ is a constant parameter with dimensions of inverse mass squared (or, equivalently, squared
length). We have
\begin{equation}
    f'(R) = 1 + 2 \beta R \,,
\end{equation}
and the trace of the metric field equations \eqref{eomg} yield
\begin{equation}
    R = - \kappa T \,.
\end{equation}
Expanding the left-hand side of the latter by exploiting the FLRW decomposition of the general affine connection and also recalling, from \eqref{metrenmomform}, that $T=-\rho +3 p$, we find
\begin{equation}
    3\Big[(\dot{X}-\dot{Y})+3 H (X-Y) +(X+Y)(Z+V)-2 X Y - 2 W^{2}+2 \dot{H}+4 H^{2} \Big]=\kappa (\rho -3 p) \,. \label{trA}
\end{equation}
Furthermore, we know that the acceleration (also known as Raychaudhuri) equation for non-Riemannian geometries in its general form is \eqref{acceleq}.
Then, contracting the metric field equations with $u^{\mu}u^{\nu}$, we find
\beq
{R_{\mu\nu}u^{\mu}u^{\nu}=\frac{\kappa}{f'}\Big( T_{\mu\nu}u^{\mu}u^{\nu}+\frac{T}{4}\Big) +\frac{\kappa T}{4}} \,.
\eeq
The above if fairly general. \\
Given that the energy-momentum tensor has the usual perfect fluid form, we get
\begin{equation}
    {R_{\mu\nu}u^{\mu}u^{\nu}= \frac{\kappa}{4}\left[\frac{3}{(1-2 \beta \kappa T)} (\rho + p)+ (-\rho + 3 p) \right] }
\end{equation}
and, substituting the latter into the above acceleration equation, it follows that
\begin{align}
\frac{\ddot{a}}{a}&=- {\frac{\kappa}{12} \left[\frac{3}{(1-2 \beta \kappa T)} (\rho + p)+ (-\rho + 3 p) \right]} \nonumber \\
&+2\left( \frac{\dot{a}}{a} \right)\Phi +2\dot{\Phi}
+\left( \frac{\dot{a}}{a} \right)\left(A+\frac{C}{2}\right) +\frac{\dot{A}}{2}-\frac{A^{2}}{4}-\frac{1}{4}AC 
-A\Phi-C \Phi \,.
\end{align}
In the above, the terms on the first line of the right-hand side represent the effect of the energy-momentum part to the acceleration. Of course we see that for $\beta=0$ we get the usual contribution $-\kappa/[6(\rho+3 p)]$, which always decelerates the expansion. Here, however, we see that this part could just as well speed up the expansion as long as $1- 2 \beta \kappa T<0$ {and $-\rho + 3 p<0$}. The terms on the second line of the above acceleration equation are the contributions of the hypermomentum part of the hyperfluid. This can be seen in a clearer way by employing the relations \eqref{Beq}-\eqref{Phieq}. \\
Now, using the acceleration equation, we may eliminate the term $\dot{H}=\frac{\ddot{a}}{a}-H^2$ from the trace equation \eqref{trA} and derive the modified first Friedmann equation, which reads
\begin{align}
H^{2} & =\frac{\kappa}{{6}}(\rho- 3 p)-\frac{{1}}{2}\Big[(\dot{X}-\dot{Y})+3 H (X-Y) +(X+Y)(Z+V)-2 X Y - 2 W^{2} \Big] \nonumber \\
& + {\frac{\kappa}{12} \left[\frac{3}{(1-2 \beta \kappa T)} (\rho + p)+ (-\rho + 3 p) \right]} \nonumber \\
& {-2\left( \frac{\dot{a}}{a} \right)\Phi -2\dot{\Phi}
-\left( \frac{\dot{a}}{a} \right)\left(A+\frac{C}{2}\right) -\frac{\dot{A}}{2}+\frac{A^{2}}{4}+\frac{1}{4}AC 
+A\Phi+C \Phi} \,. \label{firstmodfrgen}
\end{align}

In order to better understand the cosmological aspects of the model, let us now make the following assumptions on the hypermomentum variables (recall that we have already derived $\omega=3\phi$ and the vanishing of the dilation hypermomentum):
\begin{equation}
    \phi \neq 0 \,, \quad \psi = \chi = \zeta = 0 \,,
\end{equation}
that is we are considering
\begin{equation}
    \Delta_{\alpha \mu \nu} = \phi \left( h_{\mu \alpha} u_\nu + 3 u_\alpha u_\mu u_\nu \right) \,.
\end{equation}
Therefore, we are left with
\begin{equation}\label{ABCPhiPfin}
    A = \frac{\kappa \phi}{F} \,, \quad B = - \frac{2\kappa \phi}{F} \,, \quad C = - \frac{3 \kappa \phi}{F} \,, \quad \Phi = - \frac{\kappa \phi}{2F} - \frac{\dot{F}}{4F} \,, \quad P=0 \,.
\end{equation}
Moreover, from \eqref{connvarintermsoftnm} we have
\begin{equation}\label{WVZYXfin}
    W=0 \,, \quad V = - \frac{3}{2} \frac{\kappa \phi}{F} \,, \quad Z = \frac{\kappa \phi}{2F} \,, \quad Y = - \frac{\kappa \phi}{2F} - \frac{\dot{F}}{2F} \,, \quad X = {-\frac{\kappa \phi}{2F}} + \frac{\dot{F}}{2F} \,.
\end{equation}
Besides, let us consider a hypermomentum preserving hyperfluid, that is
\begin{equation}
    \rho_c = \rho \,, \quad p_c = p \,.
\end{equation}
The latter imply that the canonical and the metric energy-momentum tensor coincide, namely
\begin{equation}
    t_{\mu \nu} = T_{\mu \nu} = \rho u_\mu u_\nu + p h_{\mu \nu} \,, \quad t = T = - \rho + 3 p \,.
\end{equation}
Under this assumption, the conservation laws of the cosmological hyperfluid become
\begin{align}
    & \hat{\nabla}_\nu \left( \sqrt{-g} {\Delta_\lambda}^{\mu \nu} \right) = 0 \,, \\
    & \tilde{\nabla}_\mu {T^\mu}_\alpha = \frac{1}{2} \Delta^{\lambda \mu \nu} R_{\lambda \mu \nu \alpha} \,.
\end{align}
Additionally, we assume that the perfect fluid variables are related through a barotropic equation of state of the usual type, namely
\begin{equation}\label{baroeq}
    p=w\rho \,,
\end{equation}
where $w$ is a barotropic index. Consequently, one can prove that the conservation laws above, once considered in the FLRW setup, yield
\begin{align}
    & \dot{\rho} + 3 H (1+w) \rho = 0 \,, \label{cons1FLRW} \\
    & \dot{\phi} + 3 H \phi = 0 \,, \label{cons2FLRW}
\end{align}
which describe the evolution of $\rho$ and $\phi$, respectively.
Eq. \eqref{cons2FLRW} can be immediately integrated to get
\begin{equation}\label{phisol}
    \phi = \phi_0 \left( \frac{a_0}{a} \right)^3 \,,
\end{equation}
where we have considered that for some fixed time $t = t_0$ we have $a(t_0) = a_0$ and $\phi(t_0) = \phi_0$. \\
We will now proceed by focusing on the weak coupling limit $|\beta \kappa T |<<1$.

\subsection{The weak coupling limit $|\beta \kappa T |<<1$}

It is interesting to study what happens when the quantity $|\beta\kappa T|$ is small compared to the unit. In this case, we can ignore terms that are of quadratic and higher order in $|\beta \kappa T|$. We will consider matter components that are different from radiation (i.e., $T\neq 0$). \\
In the weak coupling limit, with the previously introduced assumptions, the modified first Friedmann equation becomes
\begin{equation}
    H^{2}=\frac{\kappa \rho}{3} +\frac{(1+w)}{2(-1+3 w)}\beta \kappa^{2}T^{2}-\frac{1}{2}(\dot{X}+\dot{Y})-\frac{1}{2}H\Big( 3X-Y+2Z+2V \Big) -\frac{1}{2}(X-Y)(Z+V)+XY \,,
\end{equation}
that is, using the relations in \eqref{WVZYXfin} in order to express the right-hand side entirely in terms of the sources,
\begin{equation}\label{midstepff}
    H^2 = \frac{\kappa \rho}{3} +\frac{1}{2}\beta \kappa^{2}(1+w)(-1+3 w)\rho^{2} -H\frac{\dot{F}}{F}-\frac{1}{4}\left(\frac{\dot{F}}{F}\right)^{2}+{\frac{1}{4}}\left(\frac{\kappa\phi}{F}\right)^{2}+\frac{\kappa}{{2} F} \left( \dot{\phi}+3H\phi \right) \,.
\end{equation}
Note that, remarkably, the terms containing the double derivative of $F$ have cancelled out and only first order derivatives (of all quantities) appear.
Furthermore, using the conservation law \eqref{cons2FLRW} to eliminate $\dot{\phi}$ in \eqref{midstepff}, we get
\begin{equation}
    H^2 = \frac{\kappa \rho}{3} + \frac{1}{2}\beta \kappa^{2}(1+w)(-1+3 w)\rho^{2} - H \frac{\dot{F}}{F} - \frac{1}{4} \left( \frac{\dot{F}}{F} \right)^2 + {\frac{1}{4}} \left( \frac{\kappa \phi}{F} \right)^2 \,.
\end{equation}
Finally, by moving the third and fourth terms of the right-hand side to the left, we observe the formation of a perfect square and we are left with
\begin{equation}\label{lastfrbefdust}
    \left(H+\frac{1}{2}\frac{\dot{F}}{F}\right)^{2}=\frac{\kappa \rho}{3} +\frac{1}{2}\beta \kappa^{2}(1+w)(-1+3 w)\rho^{2}+{\frac{1}{4}}\left(\frac{\kappa\phi}{F}\right)^{2} \,.
\end{equation}
We will now look for exact solutions of this cosmological model.

\subsubsection{Exact solutions}

To first order in $\beta\kappa T$ we have
\begin{equation}
\frac{\dot{F}}{2 F}\approx -\beta \kappa \dot{T} \,.
\end{equation}
If we assume $p$ and $\rho$ to be related by eq. \eqref{baroeq}, we have $\dot{T}=-3H (1+w)T$ and the left-hand side of \eqref{lastfrbefdust} boils down to
\begin{equation}
\left(H+\frac{1}{2}\frac{\dot{F}}{F}\right)^{2}\approx H^{2}[1+3 \beta \kappa T (1+w)]^{2}\approx H^{2}\Big(1+6\beta \kappa (1+w)T\Big) \,.
\end{equation}
Thus, to first order in $\beta \kappa T$ we get
\begin{equation}
    H^2 = \frac{\kappa \rho}{3} + \frac{3}{2} \beta \kappa^2 \rho^2 (1+w)(1-3w) + \frac{\kappa^2 \phi^2}{4} + \frac{1}{2} \beta \kappa^3 \rho \phi^2 (1-9 w^2) \,. \label{eqffwithw}
\end{equation}
Note the interesting coupling between $\rho$ and $\phi^{2}$ appearing on the right-hand side of the above modified Friedmann equation. This term ties together the classical perfect fluid contribution with that of the fluid microstructure (i.e., hypermomentum).  Besides, from eq. \eqref{cons1FLRW} we have 
\begin{equation}\label{rhosolw}
    \rho = \frac{c_0}{a(t)^{3(1+w)}}
\end{equation} 
and, using also eq. \eqref{phisol}, the above Friedmann equation can be rewritten as
\beq
H^{2}=\frac{\gamma_{1}^{2}}{a^{3(1+w)}}+\beta(1-3 w)\frac{\gamma_{2}^{2}}{a^{6(1+w)}}+\frac{\gamma_{3}^{2}}{a^{6}}+\beta(1-9 w^{2})\frac{\gamma_{4}^{2}}{a^{3(3+w)}} \,,
\eeq
where
\begin{equation}
    \gamma_1^2 := \frac{\kappa c_0}{3} \,, \quad \gamma_2^2 := \frac{3}{2} \kappa^2 c_0^2 (1+w) \,, \quad \gamma_3^2 := \frac{\kappa^2}{4} \phi_0^2 a_0^6 \,, \quad \gamma_4^2 := \frac{1}{2} \kappa^3 c_0 \phi_0^2 a_0^6 \,.
\end{equation}
Now, on the logical assumption that $w \in (-1,1)$, for early times the last term on the right-hand side of the latter is dominant over every other term in the same equation and we may approximate
\beq
H^{2}\approx\beta(1-9 w^{2})\frac{\gamma_{4}^{2}}{a^{3(3+w)}} \,.
\eeq
This is then integrated straightforwardly to give 
\beq
a(t) =\left[\frac{3(3+w)}{2}\gamma_{4}\sqrt{\beta(1-9w^{2})}t+C\right]^{\frac{2}{3(3+w)}}
\eeq
for $\beta (1-9w^{2})>0$, while for $\beta (1-9w^{2})<0$ there is no real solution.

We may now distinguish three particular cases, which differ in the value of $w$: $w=0$, $w=1/3$, and $w=-1$.\footnote{In the following we will directly select only real solutions.}

\paragraph{Case $w=0$:} This is the case of a dust (i.e., non-relativistic pressureless matter) dominated Universe, that is
\begin{equation}
    p=0 \,, \quad w=0 \,.
\end{equation}
Here we have that $\dot{T}=-3HT$ and, upon using the above, the left-hand side of \eqref{lastfrbefdust} becomes
\begin{equation}
\left(H+\frac{1}{2}\frac{\dot{F}}{F}\right)^{2}=H^{2}(1+3 \beta \kappa T)^{2} \,.
\end{equation}
Thus, to first order in $\beta \kappa T$ we find
\begin{equation}\label{finaleqH2}
    H^2 = \frac{\kappa \rho}{3}+\frac{3}{2}\beta \kappa^{2}\rho^{2}+\frac{\kappa^{2}\phi^{2}}{4}+\frac{1}{2}\beta \kappa^{3}\rho \phi^{2} \,.
\end{equation}
One may then consider the early stages of the Universe, that is early times, in which $a$ is relatively small. In this case, taking into account \eqref{phisol} and the fact that from \eqref{cons1FLRW} we have 
\begin{equation}\label{rhosol}
    \rho = \frac{c_0}{a(t)^3} \,,
\end{equation}
with $c_0$ constant, the term along $\rho \phi^2$ on the right-hand side of \eqref{finaleqH2}, which goes as $1/a^9$, is dominant with respect to the others, which, in turn, can be neglected. Thus, we are left with
\begin{equation}
    H^2 = \frac{1}{2} \beta \kappa^3 \rho \phi^2 \,.
\end{equation}
Solving the latter for $H$ we find
\begin{equation}
    H = \pm \frac{\kappa^{3/2}\beta^{1/2}}{\sqrt{2}} \rho^{1/2} \phi
\end{equation}
and, plugging \eqref{phisol} for $\phi$ and \eqref{rhosol} for $\rho$ into this last equation, which yields
\begin{equation}
    H = \frac{\beta_1 \phi_0 a_0^3}{a^{9/2}} \,, \quad \beta_1 := \pm \frac{\kappa^{3/2}(\beta c_0)^{1/2}}{\sqrt{2}}
\end{equation}
after integration we obtain 
\begin{equation}
    a(t) = \frac{3^{4/9}}{2^{2/9}} \left[\beta_1 \phi_0 a_0^3 (t-t_0) + c_1 \right]^{2/9} \,,
\end{equation}
where $c_1$ is a constant. \\
On the other hand, for late times, $a>>1$ and the contribution $1/a^9$ can be ignored compared to the ones that go with $1/a^3$ and $1/a^6$. In this case, we are left with
\begin{equation}
    H^2 = \frac{\kappa \rho}{3}+\frac{3}{2}\beta \kappa^{2}\rho^{2}+\frac{\kappa^{2}\phi^{2}}{4} \,,
\end{equation}
which can be solved for $H$, yielding
\begin{equation}
    H = \pm \frac{\sqrt{\kappa}}{2 \sqrt{3}} \left( 3 \kappa \phi^2 + 4 \rho + 18 \beta \kappa \rho^2 \right)^{1/2} \,.
\end{equation}
Using \eqref{phisol} and \eqref{rhosol} in this last equation, that is considering
\begin{equation}
     H = \pm \frac{\sqrt{\kappa}}{2 \sqrt{3}} \left[ \frac{(18 \beta c_0^2 \kappa + 3 a_0^6 \kappa \phi_0^2)}{a^6} + 4 \frac{c_0}{a^3} \right]^{1/2} \,,
\end{equation}
after integration we get
\begin{equation}
\begin{split}
    a(t) & = \frac{1}{2^{2/3}} \left[ 3 \left( - 6 \beta c_0 \kappa - \frac{a_0^6 \kappa \phi_0^2}{c_0} + \left( \sqrt{c_0 \kappa} (t-t_0) \pm 2 \sqrt{3} \sqrt{c}_0 c_2 \right)^2 \right) \right]^{1/3} \\
    & = \frac{1}{2^{2/3}} \left[ 3 \left( - 6 \beta c_0 \kappa - \frac{a_0^6 \kappa \phi_0^2}{c_0} + c_0 \kappa (t-t_0)^2 \pm 4 \sqrt{3} c_0 \sqrt{\kappa} (t-t_0) c_2 + 12 c_0 c_2^2 \right) \right]^{1/3} \,,
\end{split}    
\end{equation}
with $c_2$ constant. \\
Finally, in the case in which the perfect fluid characteristic $\rho$ is dominating with respect to shear hypermomentum, one might ignore all $\phi$ contributions. In this case, we have 
\begin{equation}
   H^2 = \frac{\kappa c_0}{3 a^3} + \frac{3}{2} \beta \kappa^2 \frac{c_0^2}{a^6} \,.
\end{equation}
Solving the latter equation, we get
\begin{equation}
\begin{split}
    a(t) & = \frac{1}{2^{2/3}} \left[ 3 \left( - 6 \beta c_0 \kappa + \left( \sqrt{c_0 \kappa} (t-t_0) \pm \sqrt{6} c_2 \right)^2 \right) \right]^{1/3} \\
    & = \frac{1}{2^{2/3}} \left[ 3 \left( - 6 \beta c_0 \kappa + c_0 \kappa (t-t_0)^2 \pm 2 \sqrt{6 c_0 \kappa} (t-t_0) c_2 + 6 c_2^2 \right) \right]^{1/3} \,,
\end{split}
\end{equation}
where $c_2$ is a constant. \\
Note that, in all the sub-cases in which $w=0$ discussed above, we have non-trivial contributions to $a(t)$ depending on the coupling constant $\beta$, as expected of course.

\paragraph{Case $w=1/3$:} This case should be considered separately since now we have that $T=0$, which means also that $f=0$, $F=1$, and $\dot{F}=\ddot{F}=0$. In addition, now
\begin{equation}
    \rho = \frac{c_0}{a(t)^4} \,.
\end{equation}
and the acceleration equation takes the form (using also the evolution equation of $\phi$)
\beq
\frac{\ddot{a}}{a}=-\frac{\kappa \rho}{3}-\frac{\kappa^{2}\phi^{2}}{2} \,.
\eeq
In addition, we immediately see that eq. \eqref{eqffwithw} boils down to
\begin{equation}
    H^2 = \frac{\kappa \rho}{3} + \frac{\kappa^2 \phi^2}{4} \,, 
\end{equation}
that is, in the case at hand,
\begin{equation}\label{eqw13}
    H^2 = \frac{\kappa c_0}{3a^4} + \frac{\kappa^2 \phi_0^2 a_0^6}{4 a^6} \,.
\end{equation}
Observe that there is no term along the coupling constant $\beta$, meaning that the cosmology in this case is not affected by the $\beta R^2$ term in the action. However, we still have a non-trivial hypermomentum contribution. \\
Then, if we consider early times, the second term on the right-hand side of the above equation is dominant with respect to the first one and we are left with
\begin{equation}
    H^2 = \frac{\kappa^2 \phi_0^2 a_0^6}{4 a^6} \,,
\end{equation}
which is solved by
\begin{equation}
    a(t) = \pm \sqrt{3} \left[ \kappa a_0^3 \phi_0 (t-t_0) + c_3 \right]^{1/3} \,, \label{aearly}
\end{equation}
where $c_3$ is an integration constant. This is also the case of a shear-dominated Universe (recall that, here, the purely dilation part of hypermomentum vanishes), as the $\phi$ contribution to \eqref{eqw13} is dominant with respect to the one of $\rho$. \\
On the other hand, for late times the $\rho$ contribution is dominant with respect to the one by $\phi$ and we get
\begin{equation}
    H^2 = \frac{\kappa c_0}{3a^4} \,,
\end{equation}
which yields
\begin{equation}
    a(t) \propto t^{1/2} \,. \label{alate}
\end{equation}
Thus, in this case the usual result of general relativity for a radiation-dominated Universe is recovered. \\
It is also possible to find exact parametric solutions to eq. \eqref{eqw13}. Indeed, introducing a parameter $\theta>0$, through
\beq
a(\theta)=\sqrt{\frac{\gamma_{1}}{\gamma_{2}}} \sinh{\theta}
\eeq
we can then integrate \eqref{eqw13} trivially, to find
\beq
t(\theta)=- \frac{\gamma_{2}^{2}}{2 \gamma_{1}^{3}}\Big( \theta-\frac{1}{2}\sinh{2\theta} \Big)+C \label{t} \,,
\eeq
where we have abbreviated $\gamma_{1}^{2}=\kappa c_{0}/3$ and $\gamma_{2}^{2}=\kappa^{2}\phi_{0}^{2}a_{0}^{6}/4$ and the sign in \eqref{t} is chosen appropriately to ensure that we are in the branch $t>0$. The latter two equations describe parametrically the evolution of the scale factor and, as expected, for early and late times reproduce the solutions \eqref{aearly} and \eqref{alate}, respectively.

\paragraph{Case $w=-1$:} We are in the presence of cosmic inflation with
\begin{equation}
    \rho = c_0 = \text{constant} \,.
\end{equation}
Then, eq. \eqref{eqffwithw} yields
\begin{equation}
    H^2 = \frac{\kappa c_0}{3} + \frac{\kappa^2(1-16 \beta \kappa c_0)}{4} \phi^2 \,. 
\end{equation}
Using \eqref{phisol} and integrating, in the case at hand we get
\begin{equation}
    a(t) = \frac{1}{(2 c_0 \kappa)^{1/3}} \left[ e^{\sqrt{c_0 \kappa} \left(\pm \sqrt{3} (t-t_0) + 3 c_4 \right)} - \frac{3}{4} \kappa^2 a_0^6 c_0 (1-16 \beta \kappa c_0) e^{\sqrt{c_0 \kappa} \left( \mp\sqrt{3} (t-t_0) - 3 c_4 \right)} \kappa \phi_0^2 \right]^{1/3} \,,
\end{equation}
with $c_4$ an integration constant. Here we have a non-trivial contribution to $a(t)$ depending on the coupling constant $\beta$.

\section{Conclusions}\label{conclusions}

In this work we have analyzed the cosmological aspects of metric-affine $f(R)$ gravity with hyperfluid. We have first derived the equations of motion of the theory by varying the action with respect to the metric and the independent connection. Then, considering a FLRW background, we have derived the modified Friedmann equations of the model in the presence of a perfect cosmological hyperfluid. Consequently, we have studied the $f(R)=R+\beta R^2$ case, considering purely shear hypermomentum. By analyzing the weak coupling limit in the case of hypermomentum preserving hyperfluid ($p_c=p$, $\rho_c=0$), we have found exact solutions.

More specifically, we have obtained a general solution for the case in which the usual barotropic equation $p=w \rho$ holds for the fluid, finding that the evolution of the scale factor $a(t)$ depends on the value of the barotropic parameter $w$ (and of the coupling constant $\beta$). Hence, we have then focused on the particular cases $w=0$ (dust), $w=1/3$ (radiation), and $w=-1$ (cosmic inflation with $\rho$ constant), always considering first order in $\beta \kappa T$, where $T$ is the trace of the energy-momentum tensor. Exact solutions for the cases $w=0$ and $w=1/3$ have been obtained for early and late times, taking into account the respective dominating terms in the modified first Friedmann equation. In particular, in the $w=1/3$ case, we have $T=0$ (conformally invariant matter), which can be interpreted as yielding $f=0$, $f'=F=1$, and $\dot{F}=\ddot{F}=0$. We have found that, in this case, the cosmology is not affected by the $\beta R^2$ term in the action. However, for early times the evolution of the scale factor is driven by the non-trivial shear hypermomentum variable $\phi$, while for late times the usual result of general relativity for a radiation-dominated Universe is recovered. For $w=1/3$ we have also provided an exact parametric solution, which, considered at early and late times, reproduces the respective early and late times solutions. On the other hand, in both the $w=0$ and the $w=-1$ cases we have contributions to $a(t)$ depending on the coupling constant $\beta$. Let us finally remark that, regarding the general solution depending on $\beta$ and $w$, we have found that $\beta (1-9w^2)>0$ is required in order to have real solution, while for $\beta (1-9w^2)<0$ there is no real solution.

A future investigation may be devoted to the full cosmological analysis of $f(R)$ models involving higher powers of $R$, including the derivation of exact solutions. Moreover, it would be interesting to study the possible effects induced by the presence of spin hypermomentum, which is typically associated with spacetime torsion and couplings with fermions.

\section*{Acknowledgements}

The authors thank Flavio Bombacigno for discussions.
D. I. acknowledges support by the Estonian Research Council grant (SJD14). 
L. R. would like to thank the Department of Applied Science and Technology of the Polytechnic of Turin and the INFN for financial support.  
This paper is supported by the Ministry of Education and Science of the Republic of Kazakhstan, Grant No. AP14870191.

\end{document}